\documentclass[fleqn,usenatbib,onecolumn]{mnras}

\usepackage{newtxtext,newtxmath}

\usepackage[T1]{fontenc}

\DeclareRobustCommand{\VAN}[3]{#2}
\let\VANthebibliography\thebibliography
\def\thebibliography{\DeclareRobustCommand{\VAN}[3]{##3}\VANthebibliography}

\usepackage{graphicx}	
\usepackage{amsmath}	
\usepackage{mathtools}
\usepackage{orcidlink}
\usepackage[caption=false,lofdepth,lotdepth]{subfig} 

\newcommand{\be}{\begin{equation}}
\newcommand{\ee}{\end{equation}}
\newcommand{\bary}{\begin{eqnarray}}
\newcommand{\eary}{\end{eqnarray}}

\title[Duration of GRB 211211A and Long-Lived Magnetar]{Long-Duration GRB 211211A: Internal Energy Dissipation Driven by a Long-Lived Magnetar}

\author[Nissim Fraija]{
Nissim Fraija \orcidlink{0000-0002-0173-6453},$^{1}$\thanks{E-mail: nifraija@astro.unam.mx}
\\
$^{1}$Instituto de Astronom\' ia, Universidad Nacional Aut\'onoma de M\'exico, Circuito Exterior, C.U., A. Postal 70-264, 04510 M\'exico City, M\'exico\\
}

\date{Accepted XXX. Received YYY; in original form ZZZ}

\pubyear{\the\year{}}

\begin{document}
\label{firstpage}
\pagerange{\pageref{firstpage}--\pageref{lastpage}}
\maketitle

\begin{abstract}
The most promising candidate for short-duration gamma-ray bursts (GRBs) is the merger of two neutron stars (NSs), which produces kilonovae (KNe) in the aftermath. This merging can result in a fast-spinning, highly magnetic NS, known as a millisecond magnetar, whose accretion processes can explain different phases in GRBs.    The identification of a KN associated with the atypical long-duration GRB 211211A contradicted the classification schemes of the GRB progenitors.  This study presents a comprehensive analysis of gamma- and X-ray observations, focusing on modeling X-ray data from a long-lived magnetar with two distinct fallback accretion rates ($\dot{M}\propto t^0$ and $\propto t^{\frac12}$) during the initial phase.  The internal energy dissipation of the magnetar spin-down power, through the magnetization parameter, accounts for the long duration of the prompt gamma-ray episode observed in GRB 211211A.   Furthermore, we provide a satisfactory explanation for the precursor and extended emissions following the prompt episode within the magnetar model with two fallback accretion rates. Although these accretion rates explain different characteristics,  the model that incorporates a variable accretion rate offers a more accurate description.  The current scenario for the GRB 211211A observations aligns with a compact binary merger that produces a long-lived magnetar instead of an immediate black hole.
\end{abstract}




\section{Introduction}

Extremely brief explosions known as gamma-ray bursts (GRBs) are often historically divided into two categories \citep{Kouveliotou1993}, long and short, depending on the duration of the gamma-ray non-repeating flashes called as prompt emission ($T_{90}$).\footnote{$T_{90}$ corresponds to the interval during which a burst releases between $5\%$ and $95\%$ of the total observed counts from its emission.}  Long GRBs (lGRBs) with a duration $T_{90} \gtrsim 2\,{\rm s}$ have been associated to the core collapse of massive stars leading to supernovae (SNe) \citep{1993ApJ...405..273W, 1998Natur.395..670G, 1999Natur.401..453B, 2001ApJ...550..410M, 2003Natur.423..847H, 2006ARA&A..44..507W, 2012grbu.book..169H},\footnote{Ultra-long GRBs (ULGRBs) have a duration of thousands of seconds or more, making them a subset of LGRBs \citep{Gendre2013, levan2014, Piro2014, Greiner2015Natur,Kann2018A&A}.}  and short GRBs (sGRBs) with $T_{90} \lesssim 2\,{\rm s}$ to the merger of binary compact objects \citep{1992ApJ...395L..83N}, a black hole (BH) - neutron star (NS) \citep{1992ApJ...395L..83N} or a  NS - NS, leading to kilonovae (KNe) \citep{1992ApJ...392L...9D, 1992Natur.357..472U, 1994MNRAS.270..480T, 2011MNRAS.413.2031M}.  About once every year, giant flares from an extragalactic fast-spinning highly-magnetized NS named "magnetar" could mimic sGRBs \citep{burns2021identification, Roberts+21MGF}.  In certain situations, the distributions of both short and long classes overlap in terms of $T_{90}$, and other secondary parameters, such as the hardness ratio (HR) and/or spectral lag, among others, have to be required to lead to a reasonable categorization \citep[e.g., see][]{Paciesas+99cat, Zhang+09typeI, vonKienlin+20GBM10yrcat, kann2011,  2015PhR...561....1K}.  For instance, sGRBs typically have harder spectra while lGRBs typically have softer one \citep{2013FrPhy...8..661G}.\\  

Internal shocks and/or magnetic reconnections are the most likely causes of prompt gamma-ray emission \citep{2000ApJ...537..810W, 2003ApJ...596.1080V, 2003ApJ...596.1104V}. This occurs when a significant amount of kinetic and/or magnetic energy is converted into radiation. Dissipated magnetic energy may be used to accelerate baryons in the outflow or turned into internal energy, depending on the dissipation process.  Magnetic reconnections are efficient in relativistic outflows with high magnetization \citep{2001A&A...369..694S}, whereas internal shocks are efficient in outflows with low magnetization \citep{2011ApJ...726...75S,2013ApJ...771...54S}.  Following the prompt gamma-ray episode, the afterglow phase is observed since radio wavelengths up to GeV/TeV gamma-rays.  During this phase,  the relativistic jet enters contact with the circumburst medium, transferring a lot of their energy and forming a forward shock \citep{1995ApJ...455L.143S, 1998ApJ...497L..17S, 1999ApJ...513..669K} and a reverse shock \citep{2000ApJ...545..807K, 2003ApJ...597..455K, 2016ApJ...818..190F}.   During both shocks, electrons are accelerated and cooled down by synchrotron photons up to hundreds of MeV \citep{1998ApJ...497L..17S,2009MNRAS.400L..75K, 2010MNRAS.409..226K,2013ApJ...763...71A, 2016ApJ...818..190F} and synchrotron self-Compton (SSC) processes scattering them at GeV/TeV energies \citep{2019ApJ...885...29F, 2019ApJ...883..162F,2021ApJ...918...12F, 2017ApJ...848...94F,2019arXiv191109862Z}.\\

%
%
At 13:09:59 UT on December 11, 2021, the extremely bright burst GRB 211211A triggered the Burst Alert Telescope (BAT) onboard {\it Neil Gehrels Swift Observatory} \citep{2021GCN.31202....1D} and the Gamma Burst Monitor (GBM) instrument onboard the {\itshape Fermi} satellite \citep{2021GCN.31210....1M, 2023ApJ...954L...5V}.  The duration and hardness ratio of the prompt episode were $T_{90}=34.3\pm0.6\,{\rm s}$, and $HR=0.850\pm0.015$, respectively \citep{2023ApJ...954L...5V}.\footnote{The duration was estimated in the 50-300 keV range, and the hardness ratio between 50-300 keV and 10 - 50 keV.} This burst was independently followed by the X-Ray Telescope (XRT) and Ultraviolet/Optical Telescope (UVOT) instruments aboard on Swift satellite and by the INTErnational Gamma-Ray Astrophysics Laboratory (INTEGRAL) and the CALorimetric Electron Telescope (CALET). Optical analysis located to GRB 211211A close to a nearby galaxy, SDSS J140910.47+275320.8, at a distance of $z=0.076$ (350 Mpc). 

In this paper, we derive, analyze, and model the gamma- and X-ray observations of GRB 211211A in the millisecond magnetar scenario with fallback accretion.  Considering two distinct accretion rates during the initial phase,  we show that a millisecond magnetar can account for the features of prompt emission, the precursor, and the extended gamma-ray emission.  This paper is structured as follows. In Section \ref{sec_2}, we present gamma- and X-ray observations and/or data reduction.   In Section \ref{sec_3}, we introduce the millisecond magnetar scenario considering two distinct fallback accretion rates during the initial phase.  Furthermore, we show the modeling of the X-ray light curve in the magnetar scenario, interpret the prompt gamma-ray episode through the evolution of the magnetization parameter, and discuss the results. Finally, in Section \ref{sec_5} we provide a summary.

\section{GRB 211211A: Gamma- and X-ray Observations and Data Analysis}\label{sec_2}

\subsection{Gamma- and X-ray Observations}

\subsubsection{Fermi-GBM Data}

GRB 211211A triggered the GBM instrument at 13:09:59.65 UT on 11 December 2021 \citep{2021GCN.31210....1M}.  The GBM light curve of GRB 211211A was derived, analyzed and discussed in detail in \cite{2023ApJ...954L...5V}. A time-integrated spectral fit over 0--52 s with a Band function yields $\alpha = -1.180^{+0.005}_{-0.006}$, $\beta = -2.13^{+0.02}_{-0.019}$, and $E_{\rm peak} = 545^{+12.3}_{-10.8}\,{\rm keV}$. The isotropic energy in gamma-rays in the $1-10^4\, {\rm keV}$ energy range was $E_{\rm iso}=(1.25\pm 0.003)\times 10^{52}\,{\rm erg}$.


\subsubsection{Swift-BAT and -XRT Data}

The BAT instrument triggered on GRB 211211A at 13:09:59 UT on December 11 2021 ($T_{\rm 0}$).   We retrieve BAT data from the publicly available database at the official Swift website.\footnote{\url{https://gcn.gsfc.nasa.gov/notices\_s/1088940/BA/}}   Figure (\ref{fig1c}) shows the Swift-BAT light curve at different bands: 15 - 25, 25 - 50, 50-100 and 100 - 350 keV.  The time-averaged spectral analysis in the 15 - 150 keV energy range performed by the BAT Team led to the best-fit values of the photon indexes $\Gamma_{\rm BAT}=1.515\pm 0.020$ and $1.442\pm 0.075$ for a power-law (PL) and Comptonized model, respectively.\footnote{\url{https://swift.gsfc.nasa.gov/results/BATbursts/1088940/bascript/top.html}}


We retrieve data sets available from the official Swift website that were gathered by the XRT instrument on board the Swift satellite.\footnote{\url{ https://www.swift.ac.uk/burst\_analyser/01088940/}}
The {\itshape Swift}/XRT instrument began observing GRB 211211A at 13:11:18.7 UT, $T_0=79.2$ s.  This instrument monitored GRB 211211A in the windowed-timing (WT) mode with a spectrum exposure of 214 s and the Photon Counting (PC) mode with a spectrum exposure of 3.5 ks. The best-fitting absorption column (intrinsic) is $2.8^{+0.8}_{-0.8}\times 10^{20}\,{\rm cm^{-2}}$ for the WT mode. Spectral analyzes of the XRT flux performed by the Swift team\footnote{\url{https://www.swift.ac.uk/xrt_spectra/01088940/}} led to a best-fit value of $\Gamma_{\rm XRT}=1.79\pm 0.03$ at $T_0+175\,{\rm s}$ .
Figure (\ref{fig1d}) presents the Swift-XRT lightcurve in the 15 to 50 keV energy band. A blow-up of the light curve is displayed between $2\times10^3$ and $2\times10^4\,{\rm s}$.

\subsection{Data Analyis}

\begin{figure}
{\centering

\subfloat[Sub a) GRB 211211A][\centering{BAT Lightcurve. From top to bottom: 15 - 25, 25 - 50, 50-100 and 100 - 350 keV.}]{
\includegraphics[scale=0.42]{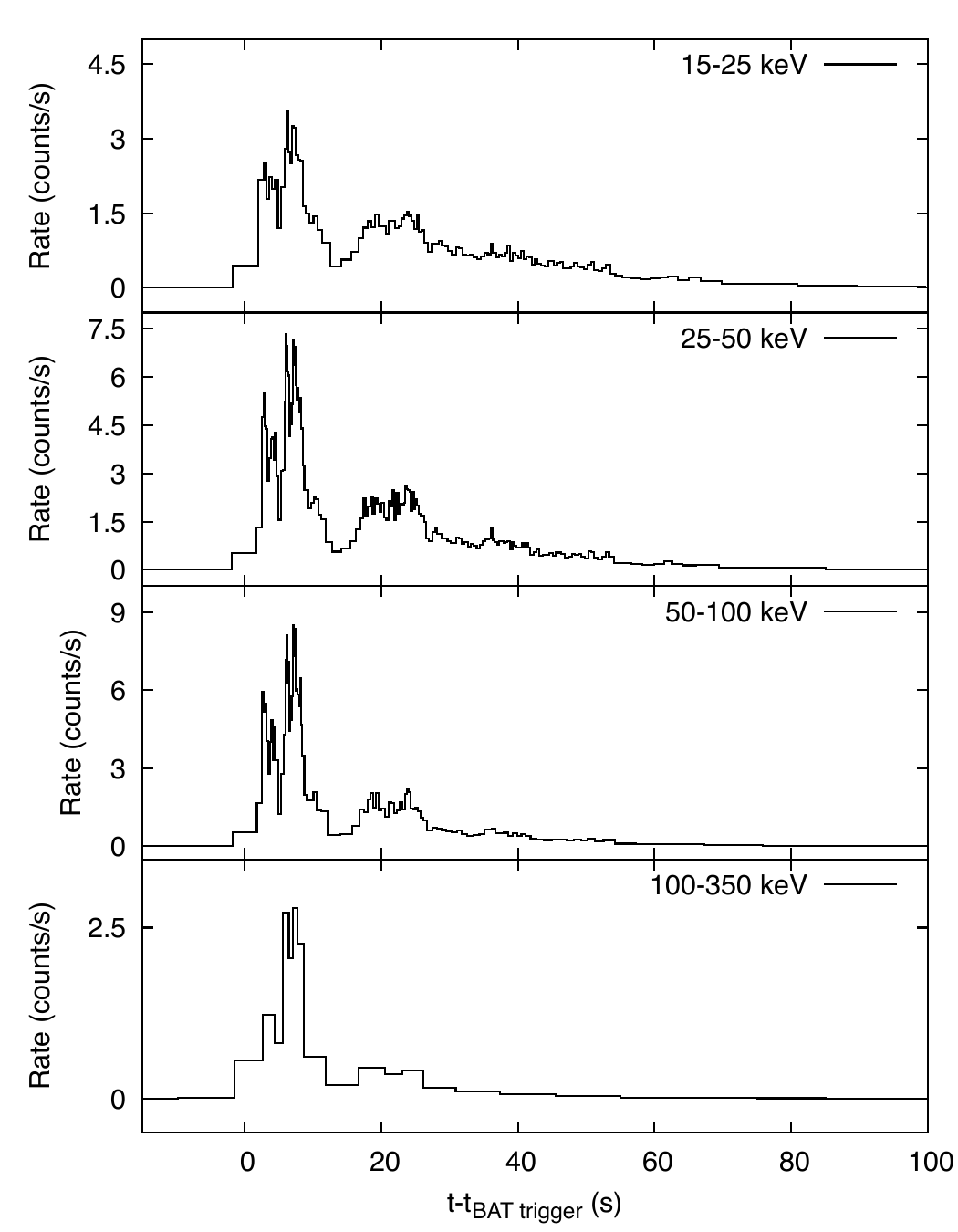}
\label{fig1c}}
\subfloat[Sub a) GRB 211211A][\centering{XRT Lightcurve. The upper subpanel shows the X-ray light curve in the 0.3 - 10 keV energy band.}]{
\includegraphics[scale=0.9]{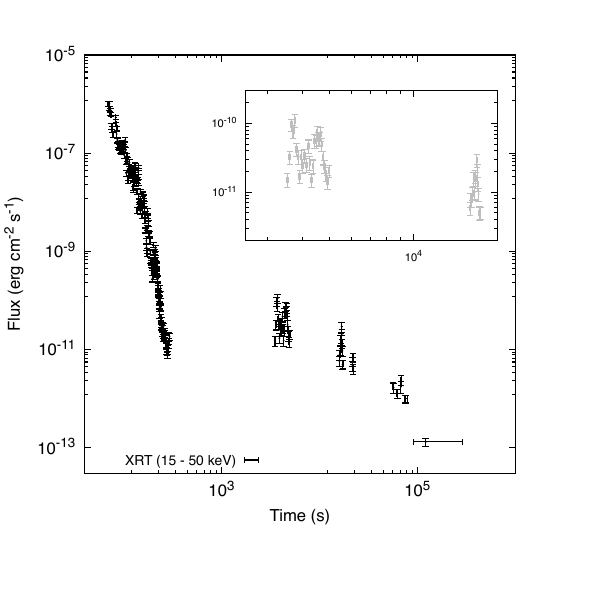}
\label{fig1d}}
\caption{Gamma-ray light curves of GRB 211211A detected by Swift-BAT and -XRT in distinct energy channels.} \label{fig1}
}
\end{figure}

The panel (\ref{fig1c}) in Figure \ref{fig1} exhibits significant gamma-ray emission in each channel.  The gamma-ray light curves exhibit three distinct components: a faint and brief gamma-ray flash known as the precursor, a period of highest activity termed the prompt episode, and a subsequent phase of late, extended emission.    The precursor lasting $0.2\,{\rm s}$, is separated $\sim 1\,{\rm s}$ from  the prompt episode. The prompt episode, the highest activity (from $\sim T_0+1\,{\rm s}$ to $T_0+13\,{\rm s}$) is formed by a series of overlapping peak structures. The late extended emission begins at $\sim T_0+13\,{\rm s}$ and is temporarily extended for almost $\sim T_0+70\,{\rm s}$.
\cite{2023ApJ...954L...5V} reported on the evolution of the variability time scale in the light curve.  They found that the prompt episode displayed a shorter timescale of variability compared to the late extended emission. The authors reported the shortest timescale variability of $2.5\,{\rm ms}$ found in the interval of 1.73 to 1.81 s.  Finally, \cite{2023ApJ...954L...5V} carried out an analysis of the extended gamma-ray emission, revealing a variability timescale that exceeds that of the prompt episode.  They fitted each spectrum with a Comptonized function and calculated the flux density at 10 keV using a broken PL function  {\small $f(t)\propto \left[ \left(\frac{t}{t_{\rm pk}}\right)^{{\rm s}\alpha_{\rm ris}}  + \left(\frac{t}{t_{\rm pk}}\right)^{{\rm s}\alpha_{\rm dec}} \right]^{-\frac{1}{s}}$} with ${\rm s}$ the smoothness parameter. The authors reported the best-fit values of the rising ($\alpha_{\rm ris}=2.0\pm 0.3$), and the decay ($\alpha_{\rm dec}=-1.5\pm 0.1$) indexes with a maximum flux at $t_{\rm pk}=20.5\pm0.9\,{\rm s}$.\\

The time-averaged spectral analysis of the Swift-BAT data in the 15–150 keV energy range yields a spectral index of $\beta = \Gamma_{\rm BAT} - 1 = 0.515 \pm 0.020$ for the PL model. Similarly, the Fermi-GBM spectral analysis below $E_{\rm peak}$ gives $\beta = \Gamma_{\rm GBM} - 1 = 0.180^{+0.005}_{-0.006}$. Finally, the spectral analysis of the Swift-XRT data in the 0.3–10 keV energy range results in $\beta = \Gamma_{\rm XRT} - 1 = 0.79 \pm 0.03$ at $T_0 + 175\,\mathrm{s}$.  It can be seen that BAT and XRT fluxes present an evolution spectral index from $\beta \approx 0.5$ to $ \approx 0.79$, respectively.
Panel (\ref{fig1d}) of Figure \ref{fig1} exhibits an initial steep decay in the XRT light curve, followed by a rising component. A notable characteristic is that tail emission reaches the minimum flux ($1.51\times 10^{-11}\,{\rm erg\,cm^{-2}\,s^{-1}}$) at 280 s, which is one order of magnitude lower than the subsequent flux ($1.93\times 10^{-10}\,{\rm erg\,cm^{-2}\,s^{-1}}$) at $3.64\times10^3\,{\rm s}$, indicating a possible injection or emission of energy from different processes.  The inset displays a series of multi-peak structures exhibiting amplitude variations spanning one order of magnitude ($\sim 10^{-11}-10^{-10}\,{\rm erg\,cm^{-2}\,s^{-1}}$) at a temporal interval from $3.1\times 10^3$ to $2.3\times 10^4$~s.

\section{Modelling, interpretation and discussion of Gamma-ray and X-ray Light Curves}\label{sec_3}

\subsection{An accreting Magnetar scenario}\label{prompt}
The millisecond magnetar stores its energy in the form of the rotational energy, which can be calculated as
\be\label{Erot}
E_{\rm}=\frac12 I\, \Omega^2\,\approx 2.6 \times 10^{52}\,{\rm erg}\,M^{\frac32}_{\rm 1.4}\,P^{-2}_{-3}\,,
\ee
where $P$ is the spin period associated with an angular frequency $\Omega=2\pi/P$ and $I\simeq 1.3\times 10^{45}\,M^{\frac32}_{\rm 1.4}\,{\rm g\,cm^2}$ \citep{2005ApJ...629..979L} is the NS moment of inertia  with $M_{\rm 1.4}=1.4\, M_\odot\,$ the NS mass.   During compact binary mergers, a substantial portion of the ejecta remains marginally bound and subsequently circularizes to form a fallback disk surrounding the remnant   \citep[e.g., see][]{2007MNRAS.376L..48R, 2007NJPh....9...17L, 2013MNRAS.435..502F}. The subsequent evolution of the disk is determined by viscous angular-momentum transport, resulting in the emergence of shallow accretion laws \citep[e.g., $\dot{M}\propto t^0$ and $\propto t^{\frac12}$;][]{2011ApJ...736..108P, 2017MNRAS.470.4925G, 2018ApJ...857...95M}.  Mass accretion following the coalescence of a NS binary is strongly dependent on the  the mass ratio of the progenitor system  \citep[e.g.,][]{2013PhRvD..87b4001H}.\\

Given the accreting mass ($M_{\rm fb}$)  over a characteristic fallback time ($t_{\rm fb}$),  the fallback accretion rate can be estimated at early and late times. Whereas a unique accretion rate is usually required at a late time, $\dot{M}\equiv \frac23\frac{M_{\rm fb}}{t_{\rm fb}}\left( \frac{\rm t}{t_{t_{\rm fb}}}\right)^{-\frac53}$ \citep{2001ApJ...550..410M, 2008ApJ...679..639Z}, two accretion rates have been considered at early times; $\dot{M}\propto t^{\frac12}$ \citep{1990ApJ...351...38C, 2011ApJ...736..108P} and $\dot{M}\propto t^0$ \citep{2017MNRAS.470.4925G, 2018ApJ...857...95M}.  Therefore, the fallback accretion rates for both scenarios can be written as \citep{2011ApJ...736..108P, 2018ApJ...857...95M}

\be\label{M_dot_co}
\dot{M}_{\rm co} \equiv  \frac23\frac{M_{\rm fb}}{t_{\rm fb}}
\begin{cases} 
\left(\frac{t}{t_{\rm fb}} \right)^{0} \hspace{0.65cm} \, {\rm t < t_{\rm fb}} \cr
\left(\frac{t}{t_{\rm fb}} \right)^{-\frac53}  \hspace{0.4cm} \, {\rm t_{\rm fb} \leq t}\hspace{0.8cm} \,, \cr
\end{cases}
\ee

and

\be\label{M_dot_va}
\dot{M}_{\rm va} \equiv  \frac23\frac{M_{\rm fb}}{t_{\rm fb}}
\begin{cases}  
\left(\frac{t}{t_{\rm fb}} \right)^{\frac12} \hspace{0.65cm} \, {\rm t < t_{\rm fb}} \cr
\left(\frac{t}{t_{\rm fb}} \right)^{-\frac53}  \hspace{0.4cm} \, {\rm t_{\rm fb} \leq t}\hspace{0.8cm} \,, \cr
\end{cases}
\ee

where $\dot{M}_{\rm co}$ and $\dot{M}_{\rm va}$ refer to scenarios in which the accretion rate at early times is constant and variable, respectively.

A constant accretion rate ($\dot{M}_{\rm co} \propto t^{0}$) can arise in a quasi-steady accretion regime, where the mass supply and viscous transport remain approximately constant over the relevant timescale.  A constant accretion rate ($\dot{M} \propto t^{0}$) implies a steady energy injection into the outflow, naturally reproducing the plateau phases observed in X-ray light curves. This behavior is consistent with the shallow decay segments frequently detected in Swift-XRT observations \citep[e.g., see][]{2006ApJ...642..354Z,2006ApJ...642..389N, 2009MNRAS.396.2038B,  Rowlinson2010}. On the other hand,  an increasing accretion rate ($\dot{M} \propto t^{1/2}$) may arise during the early phases of fallback accretion, when the inflowing material is progressively accumulated and redistributed in the disk, or when the system transitions from a propeller regime to efficient accretion. In this case, the accretion flow is not yet in steady state, leading to a gradual increase in $\dot{M}_{\rm va}$. 

Equation~\ref{M_dot_va} describes the fallback of the stellar envelope; however, this material is expected to circularize and form an accretion disk before being accreted onto the magnetar. \citet{2011ApJ...736..108P} investigated this scenario by constructing one-zone $\alpha$-disc models using the angular-momentum profiles of massive, rotating progenitors from \citet{2002RvMP...74.1015W}, as simulated with \textsc{GR1D} \citep{2010CQGra..27k4103O}. Their results show that the presence of a disk can significantly modify the temporal evolution of the accretion rate.

Once the millisecond magnetar is formed, the NS might be subject to fallback accretion.  This accretion depends on the dipole magnetic moment $\mu=BR^3$ and the  Alfv\'en radius which is given by  
\be\label{rm}
r_{\rm m}\simeq 22.3 \,{\rm km}\, M^{-\frac17}_{\rm 1.4}\,\dot{M}^{-\frac27}_{\rm w, -2}\,B^{\frac47}_{15}\,R^{\frac{12}{7}}_{\rm 6.1}\,,
\ee
with $R_{\rm 6.1}\simeq 1.2\times 10^6\, {\rm cm}\,$  the NS radius, $B$ the strength of the dipole magnetic field and ${\rm w=co}$ or ${\rm va}$ for both scenarios.  The other critical radii, that evolve with the spin period, are the co-rotation ($r_{\rm c}$) and the light cylinder ($r_{\rm lc}$) radii, which are 
\be\label{rm}
r_{\rm c}\simeq 17.4 \,{\rm km}\,  M^{\frac13}_{\rm 1.4} \,P^{\frac{2}{3}}_{-3}\,,
\ee
and
\be\label{rm}
r_{\rm lc}\simeq 48.5\,{\rm km}\, P_{-3}\,,
\ee
respectively.   The spin evolution of an accreting magnetar as a function of spin-down torque and accretion is given by 
\be\label{dif_eq}
I\frac{d\Omega}{dt}=-N_{\rm dip}+N_{\rm acc}\,.
\ee
For $r_{\rm m}> R$,  the spin evolution of the magnetar is influenced by the spin-down torque 
\be\label{N_dip}
N_{\rm dip} \simeq
\begin{cases} 
\frac{\mu^2\Omega^3}{c^3} \frac{r^2_{\rm lc}}{r^2_{\rm m}}  \hspace{0.8cm} r_{\rm m} \lesssim  r_{\rm lc}\,, \cr
\frac{\mu^2\Omega^3}{c^3}   \hspace{1.3cm} r_{\rm lc} \lesssim r_{\rm m}\,, \cr
\end{cases}
\ee
and the spin-down/up torque generated by accretion
\be\label{Nacc}
N_{\rm acc}=\dot{M}_{\rm w}(G\,M\,r_{\rm m})^\frac12\, \left[ 1-\left( \frac{r_{\rm m}}{r_{\rm c}} \right)^\frac32 \right]\,.
\ee
The torque generated by accretion corresponds to the spin up for $r_{\rm m}< r_{\rm c}$ and the spin down for the opposite case (the propeller regime). It is possible to estimate the electromagnetic spin-down luminosity by solving the differential equation (\ref{dif_eq}), which is given by

\be\label{Lsd}
L_{\rm sd} = \Omega (N_{\rm dip} - N_{\rm acc} )\,.
\ee

%
%
%

In the case of $N_{\rm acc}=0$, the equation (\ref{dif_eq}) becomes the classical spin-down magnetar equation 
\be\label{Ome1}
\frac{d\Omega}{dt}+\frac{\mu^2}{Ic^3}\Omega^3 =0\,,
\ee
which has as solution  $\Omega=\Omega_0(1+\frac{t}{t_{\rm sd}})^{-\frac12}$  with $t_{\rm sd}= \frac{c^3 I}{2\mu^2\Omega^2_0}$. Replacing the solution of Eq. (\ref{Ome1}) together with Eq. (\ref{N_dip}) in  Eq. (\ref{Lsd}), then the standard spin-down luminosity becomes
\be\label{L_sd_gen}
 L_{\rm sd}=\frac{\mu^2\Omega^4_0}{c^3}(1 + \frac{t}{t_{\rm sd}})^{-\alpha_L}\,,
 \ee
 with $\alpha_L=2$ \citep{2001ApJ...552L..35Z}.\\  
\subsubsection{Internal dissipation}

The spin-down luminosity can be converted  into isotropic X-ray flux through the efficiency in converting its spin-down energy to radiation ($\eta_{\rm x}$)  and the beaming factor of the magnetar wind ($f_b=1-\cos\theta_j$), with $\theta_j$ the half-opening angle. The  X-ray luminosity can be written as 
\be\label{Lx}
L_{\rm x}= \eta_{\rm x}\,f_b^{-1}\,L_{\rm sd} K(1+z)^{1-\beta_{\rm x}}\,,
\ee
where $K$ is the correction factor for the observed band \citep{2001AJ....121.2879B, ber13} with $\beta_{\rm x} \sim 0.8 - 1$. Similarly, the X-ray luminosity could be converted to X-ray flux  using  $F_{\rm x}\simeq \frac{L_{\rm x}}{4\pi d^2_{\rm z}}$.  The term $d_{\rm z}$ corresponds to the luminosity distance defined by $ d_{\rm z}=(1+z)\frac{c}{H_0}\int^z_0\,\frac{dz'}{\sqrt{\Omega_{\rm M}(1+z')^3+\Omega_\Lambda}} $ \citep{1972gcpa.book.....W}.  Hereafter, we assume a spatially flat universe $\Lambda$CDM model with  $H_0=67.4\,\,{\rm km\,s^{-1}\,Mpc^{-1}}$ and $\Omega_{\rm M}=1-\Omega_\Lambda=0.315$ \citep{2020A&A...641A...6P}.

\subsection{Modelling and Interpretation of gamma- and X-ray light curves}\label{sec_4}

The fit was performed using the MINUIT algorithm \citep{James:1975dr} incorporated in Python software. \cite{2021ApJ...918...12F} provides a comprehensive description of the procedure.  The evolution of the spin-down flux is shown with a fallback mass of $0.8 M_\odot$, and $\theta_j\simeq 4^\circ$ \citep[$f_{\rm b}\approx 2.43\times 10^{-3}$;][]{2022Natur.612..223R}.   Table \ref{tab1:fits} shows the best-fit values of the spin-down magnetic field, fallback accretion, the spin period, and the efficiency, for both scenarios;  constant and variable accretion rates.


\subsubsection{The gamma-ray light curve and the magnetization parameter}

The gamma-ray emission in the Poynting-flux-dominated regime will be generated by the magnetic reconnections, which could or could not induce internal shell collisions. In both cases, the magnetization parameter plays an important role. In the Internal-Collision-induced MAgnetic Reconnection and Turbulence (ICMART) model, the ejecta  needs moderate magnetization $1\lesssim\sigma \lesssim 100$ \citep{2011ApJ...726...90Z}. In some magnetic dissipation models,  the magnetization parameter is expected to be similar to the bulk Lorentz factor and lies in the critical range $100\lesssim\sigma \lesssim 3000$ \citep{2010ApJ...725.2209L, 2012MNRAS.420..483G}.

Irrespective of the model, the magnetization parameter is defined by
\be\label{sigma}
\sigma=\frac{L_j}{\dot{M}_j\,c^2}\,,
\ee
where $L_j=L_{\rm sd}$ represents the spin-down luminosity \citep{2009MNRAS.396.2038B} and $\dot{M}_j$ is the rate at which the baryon loading is ejected from the NS surface. In the case of a weakly-magnetized wind, it can be written as
\be
\dot{M}_j \simeq  \dot{M}_\nu\,f_{\rm cent}
\begin{cases} 
\frac{R_{\rm }}{2\,r_{\rm m}}\hspace{0.8cm}   r_{\rm m} \lesssim  r_{\rm lc}\ \cr
\frac{R_{\rm }}{2\,r_{\rm m}} \hspace{0.8cm} r_{\rm lc} \lesssim r_{\rm m} \cr
\end{cases}
\ee
with  $\dot{M}_\nu=\dot{M}_{\rm \nu, ob}(t)+\dot{M}_{\rm \nu, acc}(t)$, $f_{\rm cent}=e^{(\frac{P_{\rm c}}{P})^\frac32}$ and  $P_{\rm c}\simeq 2.7\,R^{\frac12}_{\rm }\, r^{-\frac12}_{\rm m}\,M^{-\frac12}_{1.4}$.

The terms $\dot{M}_{\rm \nu, ob}(t)$ and $\dot{M}_{\rm acc}(t)$ are associated with the mass loss rate due to different  sources of neutrinos. For a magnetized NS, the neutrino-driven wind is quasi-spherical, and the mass-loss rate due to neutrino ablation is given by \citep{2011MNRAS.413.2031M}

\be\label{M_nu-ob}
\dot{M}_{\rm \nu, ob}(t)=3\times10^{-4}\,(1+\frac{t}{t_{\rm kh}})^{-\frac52}\,e^{-\frac{t}{t_{\rm thin}}}\,M_{\odot}\,s^{-1}\,,
\ee
In the absence of accretion, the mass-loss rate declines sharply once the NS becomes optically thin to neutrinos, on a timescale $t_{\rm thin}\approx 10 - 30\,{\rm s}$. The cooling timescale $t_{\rm kh}\approx 2\,{\rm s}$ corresponds to  Kelvin-Helmholtz  \citep{2018ApJ...857...95M}.

Accretion provides a new neutrino source through the emission generated as the material settles onto the NS via the magnetic accretion column \citep{2011ApJ...736..108P}. The corresponding neutrino luminosity, $L_{\rm \nu,acc}\approx \frac{G M \dot{M}_{\rm w}}{R_{\rm }}$ with gravitational constant $G$, irradiates the polar cap \citep{2010ApJ...718..841Z}, where the outflow is launched along open magnetic field lines. This process contributes an extra mass-loss component

\be\label{M_nu-acc}
\dot{M}_{\rm acc}(t)=1.2\times10^{-5}\,M_{\rm 1.4}\,\dot{M}_{-2}^{\frac53}\,M_{\odot}\,s^{-1}\,,
\ee

which decays on the fallback timescale rather than $t_{\rm thin}$. As a result, $\dot{M}_{\rm acc}(t)$ dominates over $\dot{M}_{\rm \nu, ob}(t)$ at sufficiently late times.

\subsubsection{Electromagnetic spin-down luminosity tracks the entire Swift-XRT data}

The end of the prompt episode, followed by the onset of the afterglow phase, is characterized by the GRB tail \citep{2006ApJ...642..354Z}.  Several authors contend that central engines do not cease activity abruptly, implying that observed GRB tails may represent the gradual decline of these engines \citep{2005MNRAS.364L..42F, 2006ApJ...642..354Z, 2009MNRAS.395..955B}. This rapid decline, which challenges the external-shock model and accounts for the delay of photons, is attributed to the high-latitude emission resulting from the curvature effect \citep{2000ApJ...541L..51K}.\\ 

\begin{figure}
{ \centering
\resizebox*{0.51\textwidth}{0.4\textheight}
{\includegraphics{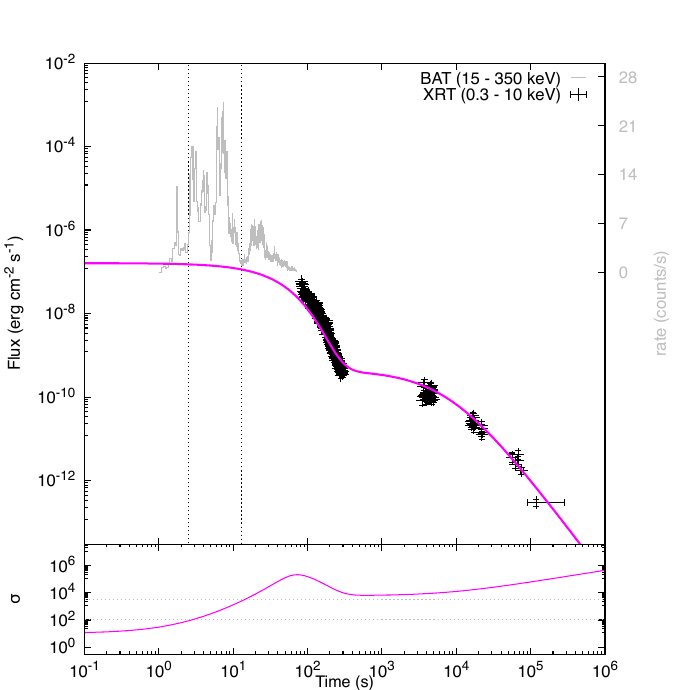}}
\resizebox*{0.51\textwidth}{0.4\textheight}
{\includegraphics{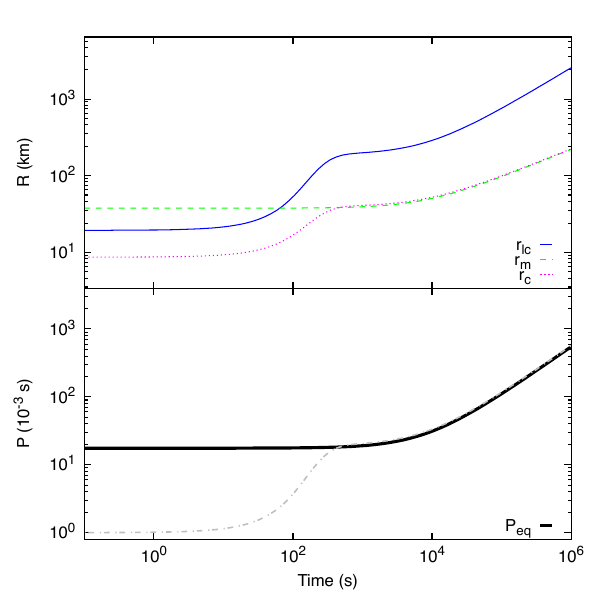}}
}
\caption{Left: The BAT light curve on counts and XRT data with the best-fit model curve generated with internal energy dissipation of the magnetar spin-down power (above) and the evolution of the magnetization parameter (below). The duration of the critical interval is marked by bold, dashed vertical lines.  Right: The evolution of the critical radii (above) and the spin period (below). These plots are obtained considering the scenario in which the accretion rate at early times is constant (Eq. \ref{M_dot_co}).  }
 \label{fig1:magnetar}
\end{figure}

\begin{figure}
{ \centering
\resizebox*{0.51\textwidth}{0.4\textheight}
{\includegraphics{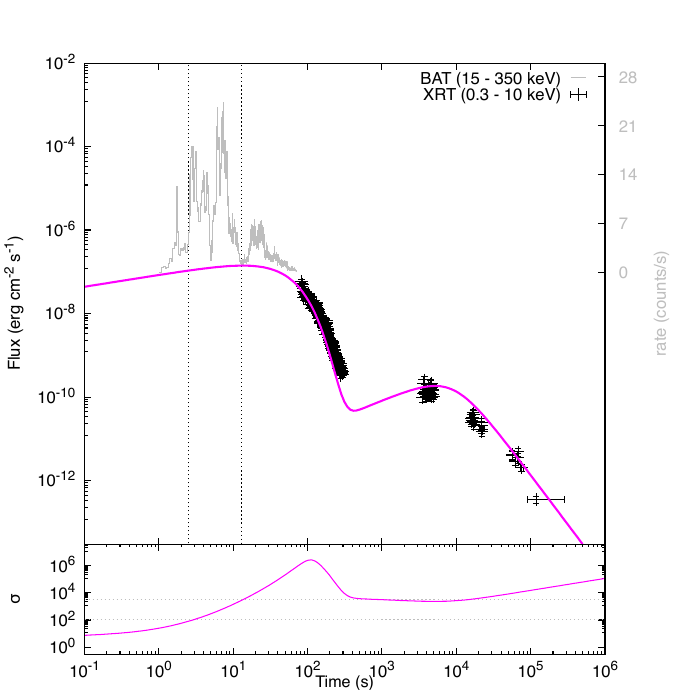}}
\resizebox*{0.51\textwidth}{0.4\textheight}
{\includegraphics{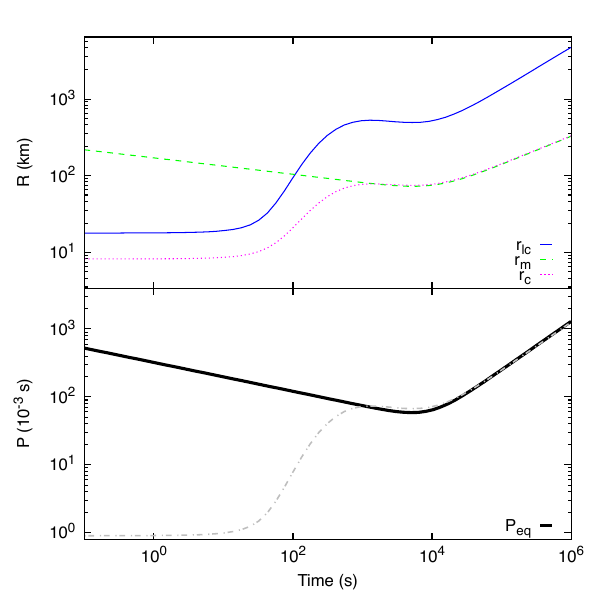}}
}
\caption{The same as Figure \ref{fig1:magnetar}, but for a variable accretion rate at early times (Eq. \ref{M_dot_va}).
}
 \label{fig2:magnetar}
\end{figure}

An analytical description of the GRB tail emission in terms of the internal energy dissipation of the magnetar spin-down power is given as follows.  For an early time much shorter than the characteristic timescale of fallback accretion ($t\ll t_{\rm fb}$),  the radius Alfv\'en evolves as ${\rm r_{m}\propto \dot{M}^{-\frac27}_{\rm w}}$, and the accretion rate as $\dot{M}_{\rm co}\propto t^0$ (Eq. \ref{M_dot_co}) and $\dot{M}_{\rm va}\propto t^{\frac12}$ (Eq. \ref{M_dot_va}). In this case, an analytic solution can be derived considering the asymptotic  limits between the critical radii $r_{\rm m}$ and $r_{\rm c}$. For $r_{\rm c}\ll r_{\rm m}$, Eq. (\ref{dif_eq}) becomes
\be
\frac{d\Omega}{dt}+\left( \frac{\mu^2}{c I r^2_{\rm m}}+\frac{\dot{M}_{\rm w}\,r^2_{\rm m}}{I} \right) \Omega = 0\,,
\ee
which has a solution $\Omega=\Omega_0\, \exp(-\frac{t}{2t_{\rm sd}})$ with $t_{\rm sd}=\frac12 \left( \frac{\mu^2}{c I r^2_{\rm m}}+\frac{\dot{M}_{\rm w}\,r^2_{\rm m}}{I} \right)^{-1}$ the characteristic spin-down timescale.  In this case, the spin-down luminosity ($L_{\rm sd}=\Omega(N_{\rm dip}-N_{\rm acc})$) is given by
\be\label{case_1}
L_{\rm sd}\simeq \left( \frac{\mu^2}{c\, r^2_{\rm m}}+ \dot{M}_{\rm w}\,r^2_{\rm m} \right) \Omega^2_0 \exp(-\frac{t}{t_{\rm sd}})\,.
\ee
For $t<t_{\rm sd}$ and an early accretion rate of $\dot{M}_{\rm co}\propto t^0$ and $\dot{M}_{\rm va}\propto t^\frac12$, the spin-down luminosity evolves as $L_{\rm sd,co}\propto t^0$ and $L_{\rm sd,va}\propto t^\frac27$, respectively, and after the spin-down luminosity decays as an exponential  $L_{\rm sd,w}\propto \exp(-t)$ with ${\rm w=co}$ and ${\rm va}$.  The internal energy dissipation of the magnetar spin-down power could describe the GRB tail emission of the prompt episode between 70 $ \lesssim t \lesssim 295\,{\rm s}$, as shown in the upper-left panels of Figures \ref{fig1:magnetar} and \ref{fig2:magnetar}.\\

The equilibrium is reached as $r_{\rm m}= r_{\rm c}$. Therefore,  the spin period is given by

\be\label{Pc}  
P_{\rm eq}\simeq 1.5\times 10^{-3}\,  {\rm s}\,\, B^{\frac67}_{15}\, R^{\frac{18}{7}}_{\rm 6.1}\, M^{-\frac57}_{\rm 1.4}\, \dot{M}^{-\frac37}_{\rm w,-2}\,,
\ee
with  $\Omega_{\rm eq}=2\pi/P_{\rm eq}$. In this regime,  the accretion term is zero  ($N_{\rm acc}(\Omega_{\rm eq})=0$), and then the spin-down luminosity is  $L_{\rm sd}=\Omega N_{\rm dip}$.   From  Eqs.  (\ref{M_dot_co}), (\ref{M_dot_va}) and (\ref{Pc}),  the spin-down luminosity $L_{\rm sd}\simeq\frac{\mu^2\,\Omega^4_{\rm eq}}{c^3}\frac{r^2_{\rm lc}(\Omega_{\rm eq}) }{r^2_{\rm c}}$ for both  fallback accretion rates is

{\small
\be\label{Lsd_eq_co}
L_{\rm sd, co} =
\begin{cases}
8.5\times 10^{46}\,{\rm \frac{erg}{s}}\,M^{\frac{17}{7}}_{\rm 1.4}B^{\frac{20}{7}}_{15}\,R^{\frac{60}{7}}_{\rm 6.1}\left(\frac{t_{3.5}}{t_{\rm fb,4}} \right)^{0} \hspace{0.84cm} \, {\rm t<t_{\rm fb}} \cr
5.6\times 10^{45}\,{\rm \frac{erg}{s}}\, B^{-\frac67}_{15}\,M^{\frac{12}{7}}_{\rm 1.4}\,R^{-\frac{18}{7}}_{\rm 6.1}\left(\frac{t_{4.5}}{t_{\rm fb,4}} \right)^{-\frac{50}{21}}  \hspace{0.1cm} \, {\rm t_{\rm fb} \leq t}, \cr
\end{cases}
\ee
}
and
{\small
\be\label{Lsd_eq_va}
L_{\rm sd, va} =
\begin{cases}
6.2\times 10^{46}\,{\rm \frac{erg}{s}}\,M^{\frac{17}{7}}_{\rm 1.4}B^{\frac{20}{7}}_{15}\,R^{\frac{60}{7}}_{\rm 6.1}\left(\frac{t_{3.5}}{t_{\rm fb,4}} \right)^{\frac57} \hspace{0.84cm} \, {\rm t<t_{\rm fb}} \cr
9.6\times 10^{45}\,{\rm \frac{erg}{s}}\, B^{-\frac67}_{15}\,M^{\frac{12}{7}}_{\rm 1.4}\,R^{-\frac{18}{7}}_{\rm 6.1}\left(\frac{t_{4.5}}{t_{\rm fb,4}} \right)^{-\frac{50}{21}}  \hspace{0.1cm} \, {\rm t_{\rm fb} \leq t}. \cr
\end{cases}
\ee
}

Due to accretion during the late phase, the spin-down luminosity evolves as $\propto t^{-\frac{50}{21}}$ rather than $\propto t^{-2}$, which would be expected in the absence of accretion. If the accretion rate evolves as $\dot{M} \propto t^{-\frac{4}{3}}$ \citep{2007NJPh....9...17L, 2011MNRAS.413.2031M}, then the spin-down luminosity would vary as $\propto t^{-\frac{40}{21}}$.\\

The left-hand panel of Figure \ref{fig1:magnetar} shows the Swift-XRT data track the electromagnetic spin-down luminosity. This curve is obtained considering the scenario in which the accretion rate at early times is constant (Eq. \ref{Lsd_eq_co}).  
We can note that the best-fit curve adequately describes the XRT data.  During the initial $\approx$ 20 s, the spin-down luminosity remains constant. It then rapidly declines until around $4\times 10^2\,{\rm s}$. Following this period, the spin-down luminosity stabilizes again until  $\approx 3\times 10^3\,{\rm s}$, after which it decreases once more. The right-hand panel of Figure \ref{fig1:magnetar} shows the evolution of the critical radii (above) and the spin period (below).  The spin period reaches its equilibrium at $\sim 4\times 10^2\,{\rm s}$ (on a timescale of $\sim10^{-1.3}\,t_{\rm fb}$). During the first seconds, the radius of the cylinder light is less than the radius of Alfv\'en,  so the spin-down luminosity is similar to that of the isolated magnetar.  The system lies in the propeller regime  so that the angular momentum losses decrease the total rotational energy.  We note that the Alfv\'en radius does not evolve up to it is reached by the corotation radius at $\approx 30\,{\rm km}$, indicating a critical transition state very close to the NS.

Figure \ref{fig2:magnetar} is the same as Figure \ref{fig1:magnetar}, but considering the spin-down luminosity produced by a variable accretion rate at early times (Eq. \ref{Lsd_eq_va}).  The left-hand panel of Figure \ref{fig2:magnetar} shows that the best-fit curve generated by internal dissipation is consistent with the XRT data. Within the first $\sim$ 70 s, the spin-down luminosity rises to its maximum, then rapidly declines until $\approx 4\times 10^2\,{\rm s}$. After $4\times 10^2\,{\rm s}$, the spin-down luminosity rises again until  $\approx 6\times 10^3\,{\rm s}$, then decreases again.  The right-hand panel of Figure \ref{fig1:magnetar} displays the evolution of the critical radii (above) and the spin period (below).  The spin period reaches its equilibrium at $\sim 4\times 10^2\,{\rm s}$.  For the scenario of variable accretion rate,  the Alfkén radius at the beginning  is larger than corotation radius  ($r_{\rm m} > r_{\rm co}$) (see the right-hand panel of Figure \ref{fig2:magnetar}). In this configuration, the magnetic field rotates faster than the inner accretion flow, preventing steady inflow. Direct accretion onto the magnetar is not allowed.
Matter continues to be supplied from larger radii and accumulates at the inner edge of the accretion disk. The surface density and pressure in this region gradually increase while the magnetosphere decreases.
Once the accumulated material exerts sufficient ram pressure to overcome the magnetic pressure, the magnetosphere is compressed inward, up to the Alfkén radius approaches the corotation radius at $r_{\rm m}\approx r_{\rm co}\approx 70\,{\rm km}$. The system naturally enters a magnetospheric gating regime, characterized by episodic accretion. 

The lower sub-panel of the left-hand panel in Figures \ref{fig1:magnetar} and \ref{fig2:magnetar} illustrates the evolution of the magnetization parameter for constant (Eq. \ref{Lsd_eq_co}) and variable (Eq. \ref{Lsd_eq_va}) accretion rates, respectively.  It can be seen that, while this parameter evolves in the critical range of $10^2 \leq \sigma\leq 3\times10^3$ during the interval [2.39: 12.1]~s for a constant accretion rate,  it lies in two intervals  [2.41: 12.9]~s and [$5.1\times 10^2$: $2.3\times 10^4$]~s when a variable accretion rate is considered.    In both cases, the prompt gamma-ray episode (the onset and end of the highest activity exhibited in the BAT data) coincides with the magnetization parameter crossings in the critical range.  Therefore, we argue that the highest activity displayed in the light curves could be produced by internal energy dissipation. Fallback accretion prolongs the high-activity phase in the BAT light curve.  After the magnetization parameter reaches its maximum value at $\sim 80\,{\rm s}$, it decreases but does not enter the critical range, so gamma-ray emission is unlikely if the accretion rate remains constant (panel in Fig. \ref{fig1:magnetar}). In contrast, under variable accretion rates, the parameter reaches the critical interval and gamma-ray emission is expected by internal energy dissipation. In this scenario,  the evolution of the magnetic parameter in the critical range coincides with the series of overlapping peak structures between $3.3\times10^3$ and $2.3\times 10^4\,{\rm s}$  (panel in Fig. \ref{fig2:magnetar}).

\subsubsection{The extended emission following the prompt phase}

We show that the extended emission derived at 10 keV and  described with the temporal $\alpha_{\rm ris}=2.0\pm 0.3$ and $\alpha_{\rm dec}=-1.5\pm 0.1$ indexes and the temporal break at $t_{\rm br}=20.5\pm 0.9\,{\rm s}$ \cite[see Fig. 4 in ][]{2023ApJ...954L...5V} is consistent with the spin-down luminosity powers of the SSC forward-shock model. Additionally, we consider the spectral $\beta\approx0.5$ and $\approx 0.9$ indexes reported by the Swift team, and the current magnetar scenario in which the accretion rate at early times is constant and variable.

\paragraph{Coasting phase.} In the coasting regime, the bulk Lorentz factor remains unchanged $\Gamma\propto t^{0}$.
The temporal evolution of the minimum and cooling electron Lorentz factors are $\gamma_{\rm m}\propto t^0$ and $\gamma_{\rm c}\propto t^{-1}$, respectively.  The characteristic synchrotron break frequencies and the corresponding maximum flux evolve as $\nu^{\rm syn}_{\rm m}\propto t^0$, $\nu^{\rm syn}_{\rm c}\propto t^{-2}$, and $F^{\rm syn}_{\rm max}\propto t^3$, respectively. The temporal evolution of the synchrotron light curves during the coasting phase is

{\small
\begin{eqnarray}
\label{syn_esp_win_App}
F^{\rm syn}_{\rm \nu}\propto
\begin{cases} 
t^{2}      \nu^{-\frac12}, \hspace{1.65cm}\nu^{\rm syn}_{\rm c}<\nu_<\nu^{\rm syn}_{\rm m} ,\hspace{.2cm} \cr
t^{2}\,\nu^{-\frac{p}{2}},\hspace{1.58cm} \nu^{\rm syn}_{\rm m}<\nu\,, \cr
\end{cases}
\end{eqnarray}
}

for the fast-cooling regime and 

{\small
\begin{eqnarray}
\label{syn_esp_win_App}
F^{\rm syn}_{\rm \nu}\propto
\begin{cases} 
t^{3}\,\nu^{-\frac{p-1}{2}}, \hspace{1.27cm} \nu^{\rm syn}_{\rm m}<\nu<\nu^{\rm syn}_{\rm c},\hspace{.2cm}\cr
t^{2}\,\nu^{-\frac{p}{2}},\hspace{1.58cm} \nu^{\rm syn}_{\rm c}<\nu\,, \cr
\end{cases}
\end{eqnarray}
}
for the slow-cooling regime. The SSC mechanism behaves as the high-energy extension of synchrotron radiation. Photons generated by synchrotron process may be up-scattered by the same electron population via inverse Compton scattering as
$\nu^{\rm ssc}_{\rm k}\simeq \gamma^2_{\rm k} \nu^{\rm syn}_{\rm k}$.  During the coasting phase, the characteristic SSC break frequencies and the corresponding maximum flux evolve as $\nu^{\rm ssc}_{\rm m}\propto t^0$, $\nu^{\rm ssc}_{\rm c}\propto t^{-4}$ and $F^{\rm ssc}_{\rm max}\propto t^4$, respectively. The temporal evolution of the SSC light curves for the fast and slow-cooling regime is

{\small
\begin{eqnarray}
\label{ssc_esp_win_App}
F^{\rm ssc}_{\rm \nu}\propto
\begin{cases} 
t^{2}        \nu^{-\frac12}, \hspace{1.85cm}\nu^{\rm ssc}_{\rm c}<\nu_<\nu^{\rm ssc}_{\rm m} ,\hspace{.2cm} \cr
t^{2}\,\nu^{-\frac{p}{2}},\hspace{1.79cm} \nu^{\rm ssc}_{\rm m}<\nu\,, \cr
\end{cases}
\end{eqnarray}
}
and
{\small
\begin{eqnarray}
\label{ssc_esp_win_App}
F^{\rm ssc}_{\rm \nu}\propto
\begin{cases}
t^{4}\,\nu^{-\frac{p-1}{2}}, \hspace{1.47cm} \nu^{\rm ssc}_{\rm m}<\nu<\nu^{\rm ssc}_{\rm c},\hspace{.2cm}\cr
t^{2}\,\nu^{-\frac{p}{2}},\hspace{1.79cm} \nu^{\rm ssc}_{\rm c}<\nu\,, \cr
\end{cases} 
\end{eqnarray}
}
respectively.

\paragraph{Deceleration phase.}  Energy injection from the magnetar into the GRB afterglow can refresh the forward shock. In general, the spin-down luminosity given by $L_{\rm sd}\propto L_0 t^{-q}$ with $q$ the energy injection index and $L_0$ the initial luminosity, is required to estimate the injected energy as  $E_{\rm inj}\propto \int L_{\rm sd} dt$. For $q < 1$, the standard SSC and synchrotron light curves are modified, and for $q=1$,  these light curves are recovered \citep{2000ApJ...535L..33S, 2006ApJ...642..354Z}.

For GRB 211211A, the spin-down luminosity (Eqs. \ref{case_1}) evolves as $L_{\rm sd, co}\propto t^0$ or $L_{\rm sd, va}\propto t^\frac27$ for $t<t_{\rm sd}$ and $L_{\rm sd, w}\propto \exp(-t)$ for $t_{\rm sd} < t $.  Once the relativistic jet begins to decelerate in the circumburst medium with energy injection given by the spin-down luminosity $\propto t^{-q}\, [t^0]\,(t^\frac27)$, the bulk Lorentz factor evolves as $\Gamma\propto t^{-\frac{q+2}{8}}\,[t^{-\frac{1}{4}}]\,(t^{-\frac{6}{23}})$. If the deceleration time is less than the spin-down time ($t< t_{\rm sd}$), then the bulk Lorentz factor according to the Blanford-McKee solution \citep{1976PhFl...19.1130B} can be estimated as \citep{2006ApJ...642..354Z}
\be \label{Gama}
\Gamma\approx \left(\frac{3}{32\pi m_p\,c^5} \right)^{\frac18}\,(1+z)^\frac38\,n^{-\frac18}\,\left(\eta\,f_b^{-1}\, L_{\rm sd} t_{\rm sd}^q\right)^\frac18 t^{-\frac{q+2}{8}}\,,
\ee
with ${\rm q=0}$ or ${\rm q=-\frac27}$. Otherwise, for $t_{\rm sd} < t$, the bulk Lorentz factor is calculated with ${\rm q=1}$. The equivalent kinetic energy is estimated to be $E_{\rm K}\approx \eta\,f_b^{-1}\, L_{\rm sd} t_{\rm sd}^q$. The term $\eta$ corresponds to the fraction of the initial rotational energy that is transferred to the equivalent kinetic energy of the blastwave, $n$ is the circumburst density and $m_p$ is the proton mass.  The temporal evolution of the minimum and cooling electron Lorentz factors are $\gamma_{\rm m}\propto t^{-\frac{2+q}{8}}\,[t^{-\frac{1}{4}}]\,(t^{-\frac{6}{23}})$ and $\gamma_{\rm c}\propto t^{\frac{3q-2}{8}}\,[t^{-\frac{1}{4}}]\,(t^{-\frac{10}{23}})$, respectively. The characteristic synchrotron break frequencies and the corresponding maximum flux evolve as $\nu^{\rm syn}_{\rm m}\propto t^{-\frac{2+q}{2}} \,[t^{-1}]\,(t^{-\frac{6}{7}})$, $\nu^{\rm syn}_{\rm c}\propto t^{\frac{q-2}{2}}\,[t^{-1}]\,(t^{-\frac{8}{7}})$ and $F^{\rm syn}_{\rm max}\propto t^{1-q}\, [t]\, (t^{\frac{9}{7}})$, respectively. During the deceleration phase,  the temporal evolution of the synchrotron light curves is \citep{2000ApJ...535L..33S, 2006ApJ...642..354Z}

{\small
\begin{eqnarray}\nonumber
\label{syn_esp_win_App}
F^{\rm syn}_{\rm \nu}\propto
\begin{cases} 
t^{\frac{2-3q}{4}} [t^{\frac12}](t^{\frac{4}{7}}) \nu^{-\frac12}, \hspace{2.45cm}\nu^{\rm syn}_{\rm c}<\nu_<\nu^{\rm syn}_{\rm m} \hspace{.2cm} \cr
t^{\frac{(4-2p)-(p+2)q}{4}}[t^{\frac{2 - p}{2}}](t^{\frac{8-3p}{7}})\,\nu^{-\frac{p}{2}},\hspace{0.45cm} \nu^{\rm syn}_{\rm m}<\nu\,, \cr
\end{cases}
\end{eqnarray}
}

and

{\small
\begin{eqnarray}\nonumber
\label{syn_esp_win_App}
F^{\rm syn}_{\rm \nu}\propto
\begin{cases} 
t^{\frac{(6 - 2p)-(p+3)q}{4}}
[t^{\frac{3 - p}{2}}](t^{\frac{12 - 3p}{4}})\,\nu^{\frac{1-p}{2}} \hspace{0.5cm} \nu^{\rm syn}_{\rm m}<\nu<\nu^{\rm syn}_{\rm c}\hspace{.2cm}\cr
t^{\frac{(4 - 2p)-(p+2)q}{4}}[t^{\frac{2-p}{2}}](t^{\frac{8-3p}{7}})\,\nu^{-\frac{p}{2}},\hspace{0.3cm} \nu^{\rm syn}_{\rm c}<\nu\,, \cr
\end{cases}
\end{eqnarray}
}
for the fast and slow-cooling regime, respectively.  During the deceleration regime, the characteristic SSC break frequencies and the corresponding maximum flux evolve as $\nu^{\rm ssc}_{\rm m}\propto t^{-\frac{3(2+q)}{4}}\,[t^{-\frac32}]\,(t^{-\frac{9}{7}})$,  $\nu^{\rm ssc}_{\rm c}\propto t^{\frac{5q-6}{4}}\,[t^{-\frac32}]\,(t^{-\frac{13}{7}})$ and $F^{\rm ssc}_{\rm max}\propto t^{\frac{6-5q}{4}}\,[t^{\frac32}]\,(t^{\frac{13}{7}})$, respectively. During the deceleration phase,  the temporal evolution of the SSC light curves is \citep{2014ApJ...787..168V}

{\small
\begin{eqnarray}\nonumber
\label{syn_esp_win_App}
F^{\rm ssc}_{\rm \nu}\propto
\begin{cases} 
t^{\frac{5q - 6}{8}}[t^{-\frac{3}{4}}](t^{-\frac{13}{14}})      \nu^{-\frac12}, \hspace{2.3cm}\nu^{\rm ssc}_{\rm c}<\nu_<\nu^{\rm ssc}_{\rm m} \hspace{.2cm} \cr
t^{\frac{(p-2)6+(2+3p)q}{8}}[t^{\frac{(p-2)3}{4}}](t^{\frac{9p-20}{14}})\,\nu^{-\frac{p}{2}},\hspace{0.3cm} \nu^{\rm ssc}_{\rm m}<\nu\,, \cr
\end{cases}
\end{eqnarray}
}
and
{\small
\begin{eqnarray}\nonumber
\label{syn_esp_win_App}
F^{\rm ssc}_{\rm \nu}\propto
\begin{cases} 
t^{\frac{(p-3)6+(7+3p)q}{8}}[t^{\frac{(p-3)3}{4}}](t^{\frac{9p-35}{14}})\nu^{\frac{1-p}{2}} \hspace{0.3cm} \nu^{\rm ssc}_{\rm m}<\nu<\nu^{\rm ssc}_{\rm c}\hspace{.2cm}\cr
t^{\frac{(p-2)6+(2+3p)q}{8}}[t^{\frac{(p-2)3}{4}}](t^{\frac{9p-20}{14}})\nu^{-\frac{p}{2}},\hspace{0.2cm} \nu^{\rm ssc}_{\rm c}<\nu\,, \cr
\end{cases}
\end{eqnarray}
}
for the fast and slow-cooling regime, respectively.

During the coasting phase, the evolution of SSC and synchrotron light curves ($F^{\rm k}_{\rm \nu}\propto t^{-2}\nu^{-\frac12}$) under the cooling condition $\nu^{\rm k}_{\rm c}<\nu < \nu^{\rm k}_{\rm m}$ with ${\rm k=ssc}$ or ${\rm syn}$ is consistent with the temporal and spetral indexes $\alpha_{\rm ris}=2.0\pm 0.3$ and $\beta\approx 0.5$, respectively. Clearly, we note that the temporal evolution of the SSC or synchrotron emission under the slow-cooling regime ($\nu^{\rm k}_{\rm m} < \nu^{\rm k}_{\rm c}$) is not consistent with $\alpha_{\rm ris}=2.0\pm 0.3$.  Since the spectral breaks evolve as $\nu^{\rm syn}_{\rm m}\propto t^0$ and $\nu^{\rm syn}_{\rm c}\propto t^{-2}$ for synchrotron, and  $\nu^{\rm ssc}_{\rm m}\propto t^0$ and $\nu^{\rm ssc}_{\rm c}\propto t^{-4}$ for SSC, the observed flux evolves in the fast-cooling regime ($\nu^{\rm k}_{\rm c}<\nu < \nu^{\rm k}_{\rm m}$) during the coasting phase.  Following the onset of jet deceleration, we note that the SSC and synchrotron light curves are consistent with $\alpha_{\rm dec}=-1.5\pm 0.1$ exclusively when $q=1$. It indicates that the spin-down luminosity was decaying exponentially (as described in Eq. \ref{case_1} for  $t_{\rm sd} < t$) when the deceleration phase started.

Taking into account the temporal ($\alpha_{\rm dec}=-1.5\pm 0.1$) and spectral ($\beta_{\rm }=0.79\pm 0.03$) index we analyzed the radiative process and the cooling condition.  If the flux is under the cooling condition $\nu^{\rm k}_{\rm c}<\nu_{\rm }$ with ${\rm k=ssc}$ or ${\rm syn}$, the spectral index would be $p=2\beta_{\rm }=1.58$, which is atypical. Alternatively, if the flux evolves under the cooling condition $ \nu^{\rm k}_{\rm m}<\nu_{\rm }<\nu^{\rm k}_{\rm c}$, the standard spectral index $p=2\beta_{\rm }+1=2.58$ is obtained. 

In the synchrotron forward-shock model, the  observations would evolve  $F^{\rm syn}_\nu\propto t^{-1.2}\,[t^{0.21}]\,(t^{1.07})$ for $ \nu^{\rm syn}_{\rm m}<\nu_{\rm }<\nu^{\rm ssc}_{\rm c}$.  In the SSC forward-shock model, the flux would evolve $F^{\rm ssc}_\nu\propto t^{-1.53}\,[t^{-0.32}]\,(t^{-0.84})$ for $ \nu^{\rm ssc}_{\rm m}<\nu_{\rm }<\nu^{\rm syn}_{\rm c}$, which strongly agrees with the temporal and spectral evolution after $t_{\rm pk}=20.5\pm0.9\,{\rm s}$.  Therefore, a transition from fast to slow-cooling regime occurred just around the onset of the deceleration time.   It shows that the evolution of the flux at 10 keV is consistent with the SSC forward-shock model powered by spin-down luminosity. It should be noted that this approach cannot discriminate between  both scenarios of fallback accretion rates.

Given a typical value of jet opening angle for GRB 211211A  ($\approx 4^\circ$) \citep{2022Natur.612..223R}, an efficiency $\eta\approx 0.2$ \citep{2001ApJ...552L..35Z, 2011A&A...526A.121D,
2013MNRAS.430.1061R}, and an inferred value  of circumburst density $n=10^{-4}-10^{-2}\,{\rm cm^{-3}}$ \citep{2022Natur.612..228T, 2022Natur.612..236M}, the bulk Lorentz factor given by Eq. (\ref{Gama}) becomes $\Gamma\approx526 - 936$. Similarly, the isotropic-equivalent kinetic energy becomes $E_{\rm K}=1.4\times 10^{53}\,{\rm erg}$.  \cite{2022Natur.612..236M} reported a significant detection of high-energy gamma-ray emission (in the $0.1-1~\rm GeV$ range) performed by the Fermi-LAT, starting approximately $103\,{\rm s}$ after the trigger time. The values found of the initial bulk Lorentz factor and isotropic-equivalent kinetic energy indicate that GRB 211211A shares similarities with the most powerful LAT-detected bursts \citep{2015ApJ...804..105F, 2017ApJ...848...15F, Ghirlanda+18Lorentz, 2019ApJ...878...52A}.

\subsubsection{An interpretation of the precursor}

The left-hand panels in Figures \ref{fig1:magnetar} and \ref{fig2:magnetar} show that the precursor cannot be produced by internal energy dissipation through the magnetization parameter because it does not evolve in the critical range.  However, it could be interpreted in the ICMART model, which requires a lower value of the magnetization parameter. We emphasize that this represents one possible mechanism that could account for the observed precursor.\\

Similarly to the internal shock scenario, the ICMART model shells interacting internally at a certain point will produce a broad pulse \citep{2011ApJ...726...90Z}.   For fast (f) and slow (s) shells with their respective Lorentz factors  $\Gamma_{\rm f,s}$ and masses  $m_{\rm f,s}$, will merge after collision with a bulk Lorentz factor of
\be
\Gamma_{\rm m}=\frac{\Gamma_{\rm f} m_{\rm f} + \Gamma_{\rm s}m_{\rm s} }{\frac{m_{\rm f}}{\Gamma_{\rm f}} + \frac{m_{\rm s}}{\Gamma_{\rm s}}}.
\ee
%

The post-collision Lorentz factor, $\Gamma_{\rm m}$, is expected to be similar to that obtained when assuming an initial magnetization parameter ($\sigma_{\rm ini}$) for both shells ($\Gamma_{\rm f,s}, m_{\rm f,s}$), followed by a significantly lower magnetization ($\sigma_{\rm end}$) in the merged shell \citep{2011ApJ...726...90Z}.  The energy dissipation efficiency in the inelastic collision is $\eta_\gamma\simeq \frac{1}{1+\sigma_{\rm end}}$ and the observed gamma-ray luminosity can be written as
\be
L_\gamma \approx \eta_\gamma \,\varepsilon_e\,L_{\rm w} f^{-1}_{\rm b}\,,
\ee
with $\varepsilon_e$ the electron equipartition parameter and $L_{\rm w}=L_{\rm sd}$ the mean kinetic luminosity.  It is worth noting that for $\sigma_{\rm end}\approx 1$, the radiative luminosity derived in \cite{2017MNRAS.472.3058B}  for internal shocks is recovered.  Finally, the corresponding observed peak energy is in the form
\be
E_{\rm peak}\simeq 32\,{\rm keV}\, \left(\frac{1+z}{2} \right)^{-1}\, \left(\frac{\phi(p)}{1/6} \right) L^\frac12_{\gamma,50}\,R^{-1}_{\rm 15}(\eta\varepsilon_e)^\frac32\,\sigma_{1.5},
\ee
where $\phi(p)$ is a factor of the spectral index.\\
The left-hand panel of Figures \ref{fig1:magnetar} and \ref{fig2:magnetar} show that the evolution of the magnetization parameter during the time interval of precursor lies in the range of $1\lesssim\sigma \lesssim 100$. This range agrees with the emission generated by the shell collisions (a baryonic origin).

\subsection{Discussion}

\subsubsection{Comparison with sGRB population in the magnetar scenario}

Population studies and modeling of some GRBs offer substantial evidence for the magnetar model as progenitor in GRBs \citep[e.g., see][]{Troja2007,Lyons2010,Rowlinson2010, ber13, Rowlinson2013,Rowlinson2014, lu14, Rea2015,Lu2015, li18, stratta2018, Becerra2019,2021ApJ...918...12F, Zou2022}. For instance, \cite{ber13} 
demonstrated that late-time activity characterized by plateau phases and precursor emissions observed in eight detected-\textit{Swift}/BAT GRBs with known redshifts poses challenges for explanation within conventional accretion-driven models, unless the central engine is identified as a newly formed magnetar.  \cite{lu14} examined a sample of 214 GRBs (including nine sGRBs) with established redshifts to test the consistency with the predictions of the magnetar model. The authors concluded that millisecond magnetars are likely present in a significant proportion, though not necessarily all, of GRBs.
\cite{li18} conducted a systematic analysis of the Swift/XRT light curves for 101 GRBs with known redshifts that exhibited plateau phases. The analysis revealed that 20\% of the examined GRBs are consistent with the  isotropic magnetar wind  scenario.

\begin{figure}
{ \centering
\resizebox*{0.51\textwidth}{0.4\textheight}
{\includegraphics{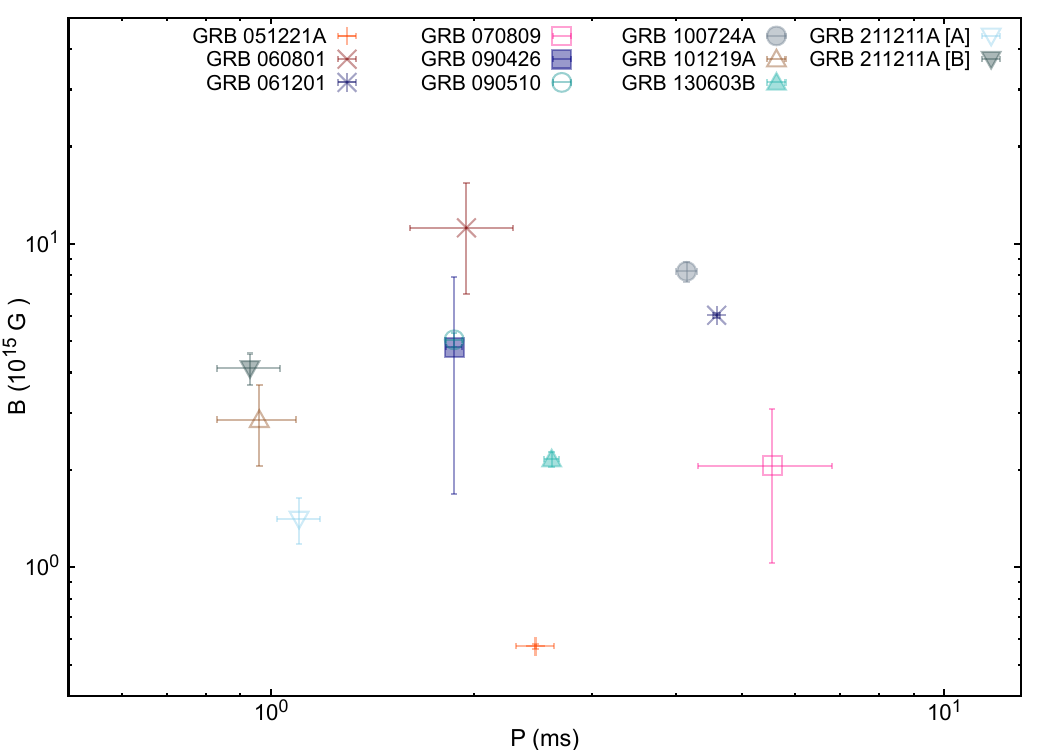}}
\resizebox*{0.51\textwidth}{0.4\textheight}
{\includegraphics{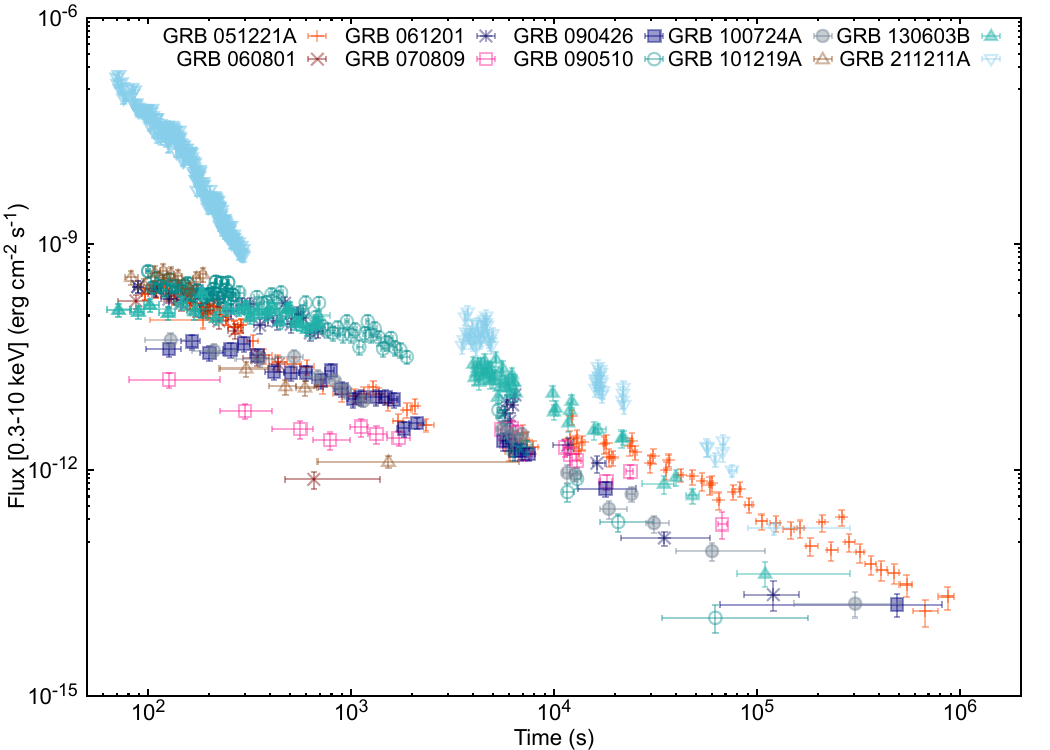}}
}
\caption{Left: Initial spin period ($P$) versus spin-down magnetic field ($B$) of a sample of sGRBs. Right: X-rays light curves of sGRBs  detected by Swift-XRT satellite, including GRB~211211A. This sample was taken from \citep{lu14}. The spin-down magnetic field and period of GRB 211211A [A] and GRB 211211A [B] correspond to the best-fit  values of constant and variable accretion scenarios, respectively.}
 \label{fig3:magnetar}
\end{figure}

The left-hand panel of Figure \ref{fig3:magnetar} shows the strength of the dipolar magnetic field and the initial spin period of the sGRB sample analyzed in \cite{lu14}, including the best-fit values found after describing GRB 211211A with both early accretion rate scenarios. This figure shows that the values found are consistent with the sGRBs distribution. The right-hand panel of Figure \ref{fig3:magnetar} compares the X-ray light curves (0.3 to 10 keV) of the sGRB sample analyzed in \cite{lu14} with that of GRB~211211A. These light curves were detected by the Swift-XRT satellite. It should be noted that the efficiency found under the assumption of a variable accretion rate falls within the range reported by \cite{2013ApJ...775...67B, lu14, li18, 2019ApJ...878...62X}. In contrast, the efficiency for the constant accretion rate scenario is comparatively lower. The X-ray observations of GRB 211211A exhibit the highest flux within this sample, thereby confirming GRB 211211A as one of the most powerful bursts. 

\subsubsection{Prior studies in GRB 211211A and the current scenario}

A KN emission was observed at optical and near-infrared (NIR) wavelengths in temporal and spatial coincidence with the burst, challenging traditional classification schemes and indicating a binary NS system as a potential progenitor, despite the duration of the prompt episode and hard spectrum \citep{2022Natur.612..228T}. Analysis indicated that the KN associated with GRB 211211A involved ejecta masses similar to those observed in other mergers but exhibited an unusually high early ultraviolet (UV) and optical brightness. \cite{2022Natur.612..232Y} explained the prompt, early X-ray observations and KN emission, and proposed a white dwarf (WD) - NS merger as the progenitor system for this event, with a post-merger magnetar providing the central engine to power the KN.  This idea of late central engine activity and the unusual progenitor was also later considered by \cite{2023ApJ...947L..21Z} and \cite{2024ApJ...971L..30H}.  \cite{2023ApJ...954L...5V} analyzed the variability timescale across the entire Fermi-GBM dataset and found that it was significantly higher during the extended gamma-ray emission than during the prompt episode. This finding linked the extended emission with the onset of the early afterglow.  Using semi-analytic radiation transport models and considering the optical and NIR observations produced by r-process nucleosynthesis, \cite{2023ApJ...947...55B} investigated this atypical burst in the collapsar scenario, specifically a low mass collapsar producing r-process elements, which are also seen in KNe.  The best-fitting models suggested that such a collapsar could explain the light curves, with parameters indicating high-velocity ejecta and an unusual distribution of ${\rm ^{56}Ni}$. However, the need for fine-tuning and extreme ejecta configurations casts doubt on this explanation. The authors argued that, while a collapsar is a possible origin for GRB 211211A, a NS merger remains the more straightforward explanation. 
 \cite{2023ApJ...957..109Z} suggested that the prompt episode of GRB 211211A may be explained by the synchrotron radiation produced by an internal shock, where the magnetic field generated by the shock decreases over time. \cite{2023ApJ...943..146C} analyzed the thermal and non-thermal emission in GRB 211211A and estimated the magnetization factor in the progenitor and the photospheric radius in a hybrid jet scenario. They found that a Poynting-flux component was present in the outflow.
\cite{2024ApJ...963..112M} showed that a photosphere model, with a structured jet, could effectively account for the long duration of GRB 211211A. \cite{2024ApJ...969...26P} performed a comparative analysis of two bursts GRB 211211A and GRB 230307A. The authors found several similarities with each other, such as the NS merger as progenitor.  \cite{2024ApJ...970....6X} observed a brief precursor lasting $\sim 0.2\,{\rm s}$ in a KN linked to GRB 211211A, highlighting its similarity to GRB 170817A. The temporal and spectral analysis suggested that this precursor was likely generated by a catastrophic flare associated with magnetoelastic or crustal oscillations of a magnetar. \cite{2025MNRAS.540.2727L} analyzed the effects of the precursor detected at $\sim 1\,{\rm s}$, considering  that the quasi-periodic oscillation (QPO) with a frequency of $\sim 20\, {\rm Hz}$ is authentic. It was shown that if the central engine is a short-lived magnetar or a hypermassive NS, the low-frequency  may be produced by instabilities inside the disk at a radius of $\sim 20 - 70\,{\rm km}$ and with a magnetic field strength of $\gtrsim 10^{13 - 14}\,{\rm G}$. \cite{2025arXiv251009744S} examined a magnetar as a remnant and established limits on the energy transferred by the magnetar's spin-down into the ejecta and circumburst medium.

The scenario proposed in this paper requires a long-lived NS with a fallback accretion rate to account for the precursor, the long duration of the prompt episode, and the extended gamma-ray emission exhibited in GRB 211211A.   Driven by the spin-down power of the millisecond magnetar, the precursor is attributed to an internal collision induced by magnetic reconnections and the atypical duration of the prompt emission to internal energy dissipation. Both component emissions are mediated by the magnetization parameter.   The extended gamma-ray emission following the prompt episode is interpreted within the early SSC afterglow scenario with energy injection during the early phase of  the fallback accretion.  We model fallback accretion using a physically motivated, piecewise prescription in which the accretion rate increases as $\dot{M}\propto t^{\frac12}$ and $\propto t^0$ during the early disk build-up phase and subsequently decays as $\propto t^{-\frac53}$ once the disk becomes isolated, ensuring angular-momentum conservation and consistency with viscous disk evolution. Although these accretion rates explain different characteristics, the model that incorporates a variable accretion
rate (i.e., $\dot{M}\propto t^{\frac12}$) offers a more accurate description.   The observed characteristics of GRB 211121A, in conjunction with the prevailing theoretical framework, suggest that the remnant resulting from the compact binary merger was a long-lived NS rather than an immediate BH.

\section{Summary}\label{sec_5}

The long-duration GRB 211211A, recognized as one of the most debated bursts due to its association with KN, has provided significant evidence for the categorization framework of GRB progenitors.  This burst, followed by a large number of satellites in gamma- and X-rays, exhibited a faint and short precursor, the prompt gamma-ray episode, and extended emission. We have proposed a scenario that requires a long-lived NS with a fallback accretion rate to account for the precursor, the prompt episode, and the extended gamma-ray emission. The precursor is attributed to an internal collision resulting from magnetic reconnection, and the atypical duration of the prompt emission is explained by internal energy dissipation.  The extended gamma-ray emission following the prompt episode is interpreted within the early SSC afterglow model, which incorporates energy injection during the initial stage of fallback accretion.
We used fallback accretion with a physically motivated, piecewise prescription. In this approach, the accretion rate increases as $\dot{M}\propto t^{\frac12}$ or $\propto t^0$ during the early disk build-up phase. Subsequently, it decays as $\propto t^{-\frac53}$ once the disk becomes isolated.  Although these accretion rates account for different features, the model that incorporates a variable accretion rate (i.e., $\dot{M}\propto t^{\frac12}$) provides a more accurate description.   The best-fit values found for the spin-down magnetic field and period are consistent with the sGRB sample analyzed in \cite{lu14}. By comparing the Swift XRT light curves in this sample, GRB 211211A corresponds to the brightest sGRBs.  The observed properties of GRB 211121A, together with the theoretical framework outlined in this paper, indicate that the remnant resulting from the compact binary merger was a long-lived NS rather than an immediate BH.

\section*{Acknowledgements}

We are grateful to the anonymous referee for valuable suggestions and detailed comments that greatly improved the quality of this work.  We thank Peter Veres and Tanmoy Laskar for helpful discussions and valuable suggestions. NF  acknowledges  financial  support  from UNAM-DGAPA-PAPIIT through the grant  IN112525.  

\section*{Data Availability}

No new data was generated or analyzed to support this research.



\bibliographystyle{mnras}
\bibliography{example} 

\newpage

\begin{table}
\centering
\caption{The best-fit values of the parameters required for modelling X-ray data of GRB 211211A. These values are derived with constant and variable accretion rates at early times.}
\label{tab1:fits}
\begin{tabular}{lcc}
\hline
\hline
Parameter & Value  & Values              \\ 
          & (constant)       & (variable)\\\hline \hline
$B\,(10^{15}\,{\rm G})$   & $1.41\pm0.23$  & $4.12\pm0.46$   \\
$t_{\rm fb}\,(10^{4}\,{\rm s})$       & $0.81\pm0.06$ & $0.93\pm0.09$ \\
$P\,({\rm ms})$       & $1.10\pm0.08$ & $0.98\pm0.10$ \\
$\eta_{\rm x}\,(10^{-3})$ & $2.30\pm0.25$ & $9.05\pm0.83$\\
\hline

\end{tabular}%
\end{table}

\appendix

\bsp	
\label{lastpage}
\end{document}